\documentclass[prb,twocolumn,superscriptaddress,showpacs,amsmath,amssymb]{revtex4}

\usepackage{graphicx}
\usepackage{latexsym}
\usepackage{amsmath}
\usepackage{amssymb}
\usepackage{amsfonts}
\usepackage{color}
\usepackage{bm}
\usepackage{verbatim}
\usepackage{color}
\begin{document}

\title{Self-consistent electronic structure of multiquantum vortices in superconductors at $T\ll T_c$}

\author{M.~A.~Silaev} \affiliation{Institute for Physics of Microstructures, Russian
Academy of Sciences, N.Novgorod, Russia}

\date{\today}

\begin{abstract}
We investigate the multiquantum vortex states in type-II
superconductor both in "clean" and "dirty" regimes defined by
impurity scattering rate. Within quasiclassical approach we
calculate self-consistently the order parameter distributions and
electronic local density of states (LDOS) profiles. In the clean
case we find the low temperature vortex core anomaly predicted
analytically in G.E. Volovik, JETP Lett. {\bf 58}, 455 (1993) and
obtain the patterns of LDOS distributions. In dirty regime the
multiquantum vortices feature a peculiar plateau in the
zero-energy LDOS profile which can be considered as an
experimental hallmark of multiquantum vortex formation in
mesoscopic superconductors.
\end{abstract}

\maketitle

\vspace{-0.2in}%

\section{Introduction}

Modern technology development provides a unique possibility to
study superconducting states at the nanoscale. Recently there has
been much experimental effort focused on the investigation of
exotic vortex states in mesoscopic superconducting samples of the
size of several coherence lengths~\cite{Mesovortices, Geim}.
 Magnetic field can
penetrate the sample in the form of a poligonlike vortex molecule
or individual vortices can merge forming multiquantum giant vortex
state with a winding number larger than unity
\cite{ExoticVortexStates}. The latter possibility is of particular
interest and the search of giant vortices in mesoscopic
superconductors was performed by means of various experimental
techniques including transport measurements \cite{Rod10,Rod11},
Bitter decoration \cite{Rod12}, magnetometry \cite{Rod13}, and
scanning Hall probe experiments \cite{Rod14}. Currently much
effort is invested to the studies of nanoscale superconducting
samples with the help of scanning tunneling microscopy (STM)
techniques \cite{RoditchevPRL2011,RoditchevPRL2009} which have
been achieved only recently and allows for the direct probe of the
structure of vortex cores through measurement of the electronic
states LDOS distribution modified by vortices.

Such STM measurements have been proven to be an effective tool of
experimental study of electronic structure of vortices in bulk
superconductors\cite{STSHess, STSmore1, STSRoditchev, STSmore2,
RMP-2007}. Indeed for the temperatures much lower than the typical
energy scale in superconductors $T\ll T_c$ the local differential
conductance of the contact between STM tip and superconductor as a
function of voltage $V$:
 \begin{equation}\label{Eq:CondLT}
  \frac{dI}{dV} (V)= \frac{dI}{dV}_{N} \frac{N({\bf r},E=eV)}{N_0}.
 \end{equation}
 where $(dI/dV)_{N}$ is a conductance of the normal metal junction and $N_0$
 is the electronic density of states at the Fermi level.
The observation of the zero-bias anomaly of tunneling conductance
at the center of singly quantized vortices\cite{STSHess,
STSRoditchev, STSmore1, STSmore2, RMP-2007} clearly confirmed the
existence of bound vortex core states predicted by Caroli, de
Gennes and Matricon (CdGM)\cite{CdGM}. In clean superconductors
for each individual vortex the energy $\varepsilon(\mu)$ of a
subgap electronic state varies from $-\Delta_0$ to
$+\Delta_0=\Delta(r=\infty)$ as one changes the angular momentum
$\mu$ defined with respect to the vortex axis. At small energies
$|\varepsilon|\ll\Delta_0$ the spectrum is a linear function of
$\mu$:
 \begin{equation}\label{Eq:SinglyQuantized}
 \varepsilon(\mu)=\omega\mu
 \end{equation}
 Here $\omega \sim \Delta_0/(k_F\xi)$ where $\xi=\hbar V_F/\Delta_0$ is coherence length,
 $k_F$ is Fermi momentum and $V_F$ is Fermi velocity.
  The wave functions of the subgap states are localized inside the vortex core because of
 the Andreev reflection of quasiparticles at the core boundary and
 determine the low energy LDOS singularity at the vortex center.

  In multiquantum vortices the spectrum of electronic states bound in the
vortex with the winding number $M$ contains $M$ anomalous branches
degenerate by electronic spin
 \cite{VolovikAnBr, multi-spectrum-num,Melnikov-Vinokur-2002,TanakaMultiquantum,SalomaaMultiquantum}:
 \begin{equation}
 \label{Volovik-spectr}
 \varepsilon_j(\mu)=\omega_j(\mu-\mu_{j})\,,
 \end{equation}
where $\omega_j \sim \Delta_0/(k_F\xi)$, index $j$ enumerates
different spectral branches ($1<j<M$),
$-k_F\xi\lesssim\mu_{j}\lesssim k_F\xi$. Each anomalous branch
intersects the Fermi level and contributes to the low-energy LDOS.
The spectrum of localized electronic states in mesoscopic
superconductors with several vortices have been shown to be very
sensitive to the mutual vortex position\cite{Silaev2009}. It has
been suggested that testing the properties of electronic spectrum
by means of the heat conductivity measurement one can directly
observe the transition to the multiquantum vortex state in
mesoscopic superconductor \cite{Silaev2008}. An alternative route
is to use STM measurement of local tunnelling conductance being
proportional to the LDOS provided $T\ll T_c$. Thus to provide the
evidence of multiquantum vortex formation revealed by STM
experiments one should find distinctive features of the order
parameter structures and LDOS profiles occurring especially in the
low temperature regime $T\ll T_c$.

Previously the low temperature properties of multiquantum vortices
have not been investigated much. The results of theoretical
studies are known only for the particular case of vortices in
clean superconductors when the electronic mean free path is much
larger than the coherence length. In this regime the contribution
of anomalous branches produces singularities of the order
parameter distribution near the vortex core in the limit $T\ll
T_c$. In particular the singly quantized vortex features an
anomalous increase of the order parameter slope at the vortex
center which is known as Kramer-Pesch effect
\cite{KramerPesch,Gygi}. The generalization to the multiquantum
vortex case was suggested in Ref.(\cite{VolovikAnomaly}) where it
was analytically predicted that doubly quantized vortex should
have square root singularity of the order parameter distribution
$\Delta=\Delta(r)$ in the limit $T\ll T_c$. Although the
structures of mutliquantum vortices have been calculated
self-consistently in the framework of Bogolubov-de Gennes theory
the vortex core anomalies have not been discussed yet
\cite{TanakaMultiquantum,SalomaaMultiquantum}. Moreover multiple
anomalous branches of electronic spectrum have been shown to
produce complicated patterns in the LDOS distributions
investigated in the framework of Bogolubov- de Gennes theory
\cite{TanakaMultiquantum,SalomaaMultiquantum}. Here we employ an
alternative approach of quasiclassical Eilenberger theory
\cite{Eilenberger} to check the predictions of vortex core
anomalies and the LDOS patterns in multiquantum vortices in clean
superconductors.

Notwithstanding the interesting physics taking place in the clean
regime the experimental realization of STM measurements of
multiquantum vortex states was implemented on Pb superconductor
\cite{RoditchevPRL2011,RoditchevPRL2009} with short mean free path
being much smaller than the coherence length. This dirty
superconductor is more adequately described within the diffusive
approximation of the electronic motion resulting in the Usadel
equations for the electronic propagators and the superconducting
order parameter \cite{Usadel}. Singly quantized vortex states in
dirty superconductors were investigated in
detail\cite{KramerJLTP74} and were shown to lack the low
temperature singularity of the $\Delta (r)$ distribution being
smoothed out by the impurity scattering of quasiparticle states.
Moreover the LDOS distribution inside vortex core does not feature
zero bias anomaly since the spectral weight of bound electronic
states is distributed smoothly between all energy scales up to the
bulk energy gap $\Delta_0$. On the other hand the multiquantum
vortex states have not been investigated in the framework of the
Usadel theory nor the LDOS distributions around multiquantum
vortices in dirty superconductors have been ever calculated.

It is the goal of the present paper to study both the
peculiarities of the multiquantum vortex structures especially at
low temperatures and the distinctive features of the electronic
LDOS near the vortices which would allow unambiguous
identification of giant vortices both in clean and dirty regimes.
This paper is organized as follows. In Sec.~\ref{theory} we give
an overview of the theoretical framework namely the quasiclassical
Eilenberger theory in clean superconductors and Usadel equation in
the dirty regime.
 We discuss the results of self-consistent calculations of the
order parameter distributions for multiquantum vortex
configurations in Sec.~\ref{op} and address the LDOS profiles in
Sec.~\ref{ldos}. We give our conclusions in Sec.~\ref{summary}.

\section{Theoretical framework} \label{theory}

\subsection{Clean limit: Eilenberger formalism}

 Within quasiclassical
approximation \cite{Eilenberger,Maki, schopohl_cm} the band
parameters characterizing the  Fermi surface is the Fermi velocity
$V_{F}$ and the density of states $N_0$. We normalize the energies
to the critical temperature $T_c$ and length to $\xi_C= \hbar
V_{F}/T_c$. The magnetic field is measured in units
$\phi_0/2\pi\xi_C^2$ where $\phi_0=2\pi\hbar c/e$ is magnetic flux
quantum. The system of Eilenberger equations for the
quasiclassical propagators $f,f^+,g$ reads
\begin{eqnarray}\label{Eq:EilenbergerF}
&{\bf n_p}\left(\nabla+i {\bf A}\right) f +
 2\omega f - 2 \Delta g=0, \\ \nonumber
 &{\bf n_p}\left(\nabla-i {\bf A}\right) f^+ -
 2\omega f^+ + 2\Delta^* g=0.
 \end{eqnarray}
 Here ${\bf A}$ is a vector potential of magnetic field, the vector ${\bf n_p}$
  parameterizes the Fermi surface and $\omega$ is a real quantity which should be taken at the
 discrete points of Matsubara frequencies $\omega_n=(2n+1)\pi T$ determined by the temperature $T$.
  The quasiclassical propagators obey
 normalization condition $g^2+ff^+=1$.
 The self-consistency equation for the gap is
  \begin{equation}\label{Eq:SelfConsClean}
  \Delta ({\bf r})=2\pi T \Lambda\sum_{n=0}^{N_d} S_F^{-1} \oint_{FS}
  f (\omega_n, {\bf r}, {\bf n_p}) d^2S_p.
 \end{equation}
 where $\Lambda$ is coupling constant, $S_F$ is a Fermi surface
 area and the integration is performed over the Fermi surface.
 Hereafter to simplify the calculations we assume the Fermi
 surface to be cylindrical and parameterized  by the angle $\theta_p$ so that ${\bf
 n_p}=(\cos\theta_p,\sin\theta_p)$. In Eq.(\ref{Eq:SelfConsClean}) $N_d (T) = \omega_d/(2\pi T)$ is a cutoff at the
   Debye energy $\omega_d$ which is expressed through physical
 parameter $T_c$ and $\Lambda$ as follows
  \begin{equation}\label{Eq:SelfConsistencyHom}
   \sum^{N_d(T_c)}_{n=0}\frac{\Lambda}{n+1/2}=1.
  \end{equation}
The LDOS is expressed through the analytical continuation of
quasiclassical Green's function to the real frequencies  \
 \begin{equation}\label{Eq:LDOSclean}
 N({\bf r})=N_0 S_F^{-1} \oint_{FS}
  Re [ g  (\omega=-iE+0, {\bf r}, {\bf n_p}) ] d^2S_p.
 \end{equation}

Assuming the vortex line to be oriented along the ${\bf z}$ axis
we choose the following ansatz of the superconducting order
parameter corresponding to axially symmetric vortex bearing $M$
quanta of vorticity $\Delta(x,y)=|\Delta|(r) e^{iM\varphi}$ where
$r=\sqrt{x^2+y^2}$ is the distance from the vortex center,
$\varphi= \arctan(y/x)$ is the polar angle. Below we neglect the
influence of the magnetic field on the vortex structure which is
justified for superconductors with large Ginzburg-Landau
parameter.

For numerical treatment of the Eqs.(\ref{Eq:EilenbergerF}) we
follow the Refs.~\cite{Maki, schopohl_cm} and introduce a Ricatti
parametrization for the propagators. The essence of this method is
a mathematical trick which allows to solve two first order Ricatti
equations instead of second-order system of Eilenberger equations.
Starting with some reasonable ansatz for the order parameter the
first order Ricatti equations are solved by the standard
procedure. Then the corrected order parameter is calculated
according to Eq. (\ref{Eq:SelfConsClean}). The badly converging
sum in Eq.(\ref{Eq:SelfConsClean}) is renormalized in a usual way
with the help of Eq.(\ref{Eq:SelfConsistencyHom}). Then one should
take into account only several terms in the sum
(\ref{Eq:SelfConsClean}). E.g. $\omega_n< 10 T_c$ is enough for
the temperature range $T>0.05 T_c$ considered at the present
paper.
 The iteration of this procedure
repeats until convergence of the order parameter is reached with
an accuracy $10^{-4} T_c$.

\subsection{Dirty limit: Usadel equations}
 In the presence of impurity scattering the
Eilenberger Eqs. (\ref{Eq:EilenbergerF}) contain an additional
diagonal self-energy term\cite{Eilenberger}. When the scattering
rate exceeds the corresponding energy gap (dirty limit) the
Eilenberger theory allows for significant simplification. In this
case the quasiclassical Usadel equations \cite{Usadel} are
applicable. The structure of singly-quantized vortices with $M=\pm
1$ in dirty superconductors was studied extensively in the
framework of the Usadel equations \cite{KramerJLTP74}
 \begin{equation}
 \omega F -\left[
 G(\mathbf{\nabla}-i\mathbf{A})^{2}F-
 F\mathbf{\nabla}^{2}G
 \right]
 =\Delta G \label{UsadelFG}
 \end{equation}
where $G$ and $F$ are normal and anomalous quasiclassical Green's
functions averaged over the Fermi surface satisfying the
normalization condition $G^2+F^*F=1$. To facilitate the analysis,
we introduce reduced variables: we use $T_{c}$ as a unit of energy
and $\xi_{D}=\sqrt{\mathcal{D}/2 T_{c}}$ where $\mathcal{D}$ is a
diffusion constant as a unit of length.  The Usadel equation is to
be supplemented with the self-consistency equation for the order
parameter
 \begin{equation}\label{Eq:SelfConsDirty}
 \Delta ({\bf r})= 2\pi T\Lambda\sum_{n=0}^{N_d} F (\omega_n,{\bf r}).
 \end{equation}
 We again neglect the influence of the magnetic field on
the vortex structure. It is convenient to introduce the vector
potential in Eq.(\ref{UsadelFG}) corresponding to a pure gauge
field which removes the phase of the order parameter
 \begin{equation}\label{Eq:VPgauge}
 {\bf A}=M \frac{{\bf z\times r}}{r^2}.
 \end{equation}
 Using $\theta-$ parametrization \cite{KramerTeta} ($F=\sin\theta$,
 $G=\cos\theta$) the Usadel equation can be rewritten in the form
 \begin{equation}
 \frac{1}{r}\frac{d}{dr} \left(r \frac{d}{dr} \theta\right)-\frac{M^{2}}{2r^2}
 \sin (2\theta)+
 \left(\Delta\cos\theta-\omega\sin\theta\right) =0. \label{Usadel-Reduced}
 \end{equation}
 Performing the renormalization of summation by $\omega_n$ in self-consistency Eq.(\ref{Eq:SelfConsDirty})
 we need to solve Eq.(\ref{Usadel-Reduced}) for a limited range of
 frequencies. We take $\omega_n\leq 10 T_c$ which allows to obtain very good accuracy. The nonlinear
 Eq.(\ref{Usadel-Reduced}) was solved iteratively. At first we
 choose a reasonable initial guess and linearize the equation
 to find the correction. The corresponding boundary problem for non-homogeneous second-order linear
 equation was solved by the sweeping method and the procedure was repeated untill convergence was reached.
 With the help of obtained solutions of Eq.(\ref{Usadel-Reduced})
 we calculated the corrected order parameter
 (\ref{Eq:SelfConsDirty}). We repeated the whole procedure to find the order parameter profile with
an accuracy $10^{-4} T_c$.

 Local density of states
 (LDOS) $N(E,r)$, which is accessible in tunneling experiments, can
 be obtained from $\theta(\omega,r)$ using analytic continuation
 \begin{equation}
 N(E,r)=Re\left[ \cos\theta ( \omega\rightarrow -i
 E+\delta,r)\right] \label{Eq:DOSdirty}
 \end{equation}
 To calculate the LDOS we solve  the
 Eq.(\ref{Usadel-Reduced}) for $\omega=-iE$. In this case it is in fact a system of two coupled second order equations
 for the real and imaginary parts of $\theta$. We use the iteration method again
 by solving repeatedly the linearized system for the corrections of
 $\theta$. The corresponding boundary problems for second-order linearized equations for ${\rm Re} \theta$ and ${\rm Im} \theta$
 were solved in turns by the sweeping method.

 \section{Order parameter structures of multiquantum vortices}
 \label{op}

\begin{figure}[h!]
 \centerline{\includegraphics[width=1.0\linewidth]{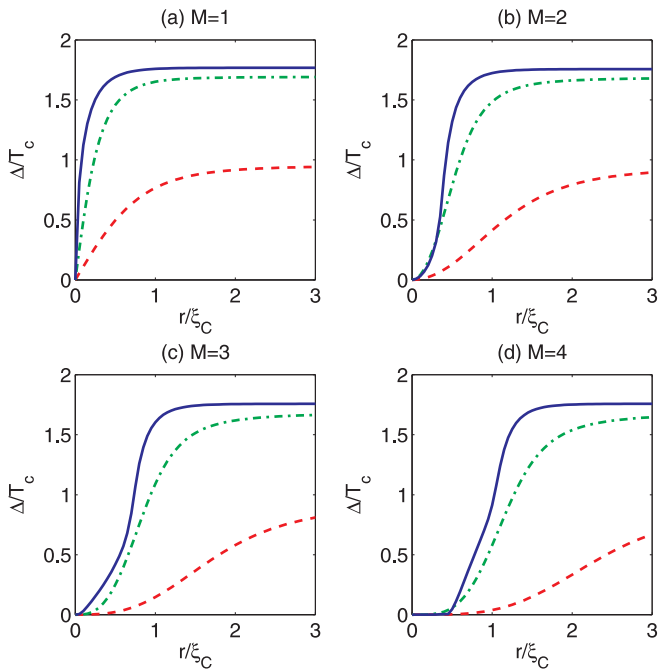}}
 \caption{\label{Fig:CleanOP}
 The distribution of the order parameter around vortex cores in
 clean superconductor at different temperatures.
 The panels (a,b,c,d) correspond to the winding numbers
 $M=1,2,3,4$. Blue solid, green dash-dotted and red dashed
 lines correspond to the temperature
 $T/T_c=0.1;\;0.5;\; 0.9$.}
 \end{figure}

 To determine the behavior of gap functions $\Delta=\Delta (r)$ in multiquantum vortices we solved numerically
 the sets of Eilenberger Eqs. (\ref{Eq:EilenbergerF},\ref{Eq:SelfConsClean}) and Usadel
 Eqs. (\ref{Eq:SelfConsDirty},\ref{Usadel-Reduced}) which describe
 the clean and dirty regimes
 correspondingly.
At first let us consider the clean regime. The order parameter
profiles in  vortices with winding numbers $M=1,\;2,\;3,\;4$ are
shown in Fig.\ref{Fig:CleanOP}(a,b,c,d) for the temperatures
$T/T_c= 0.1;\;0.5;\;0.9$. One can see that at elevated
temperatures $T=0.9 T_c$ (red dashed curves) and $T=0.5 T_c$
(green dash-dotted curve) the order parameter follows
Ginzburg-Landau asymptotic $\Delta (r) \sim r^M$ at small $r$.

\begin{figure}[h!]
 \centerline{\includegraphics[width=1.0\linewidth]{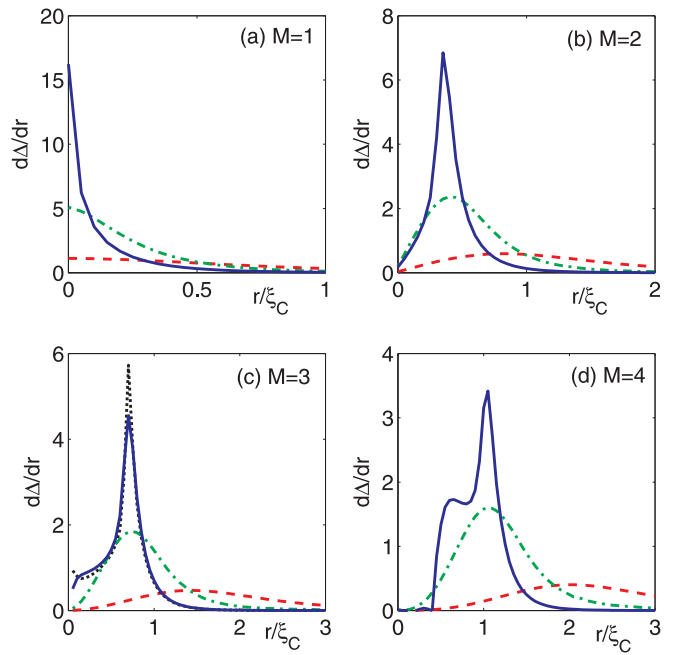}}
 \caption{\label{Fig:VortexCoreAnomaly}
 The vortex core anomaly revealed at the sharp peak of radial dependence of the order
 parameter profile derivative $d\Delta (r)/dr$
 normalized to the value $\xi_C/T_c$ around vortex cores in clean
 superconductor at different temperatures.
 The panels (a,b,c,d) correspond to the winding numbers
 $M=1,2,3,4$ correspondingly. Blue solid, green dash-dotted and red dashed
 lines correspond to the temperature
 $T/T_c=0.1;\;0.5;\; 0.9$. Dotted black line if panel (c) is for $T=0.05 T_c$; together with the blue solid curve
 in the panel (a) it demonstrates the peaked order parameter slope at the vortex center $d\Delta (r=0)/dr$
 for odd winding numbers $M$.  }
 \end{figure}

At low temperature $T=0.1 T_c$ the order parameter distribution
inside vortex core is drastically different from the
Ginzburg-Landau behaviour as shown by blue solid lines in
Fig.(\ref{Fig:CleanOP}). In particular the singly quantized vortex
in Fig.\ref{Fig:CleanOP}(a) features the Kramer-Pesch effect
\cite{KramerPesch} when the order parameter slope at $r=0$ grows
as $d\Delta/dr \sim 1/T$ when $T\rightarrow 0$. In case of
mutiquantum vortices with $M>1$ the gapless branches of electronic
spectrum (\ref{Volovik-spectr}) produce anomalies in the
multiquantum vortex core structures\cite{VolovikAnomaly}. To
observe the vortex core anomalies we plot in
Fig.(\ref{Fig:VortexCoreAnomaly}) the derivatives $d\Delta(r)/dr$
 obtained self consistently for the
vortex winding numbers $M=1,2,3,4$. In accordance with the
analytical consideration\cite{VolovikAnomaly} the vortex core
anomalies result in the singular behavior of $d\Delta(r)/dr$ at
low temperatures. We find that at $T=0.1 T_c$ in multiquantum
vortices with $M>1$ the calculated dependencies $d\Delta(r)/dr$
have sharp maxima at finite $r\neq 0$. According to the analytical
predictions these maxima originate from the square root
singularity of the order parameter which is produced by the
contribution of the anomalous energy branches of electronic
spectrum \cite{VolovikAnomaly}.

In general for higher values of winding numbers $M>1$ in the limit
$T\rightarrow 0$ one should have $M/2$ singularities of $d\Delta
(r)/dr$ for even $M$ and $(M+1)/2$ singularities for odd $M$. For
the particular examples of $M=2,4$ there are one and two peaks of
$d\Delta /dr$ at $T=0.1 T_c$ shown by blue solid line in
Fig.\ref{Fig:VortexCoreAnomaly} (b,d).
 We found that the order parameter
of $M=3$ vortex has linear asymptotic $\Delta(r)\sim r$ at small
$r$ shown in the Fig.\ref{Fig:CleanOP}(c). The slope of this
linear dependence grows at decreasing temperature which
analogously to the Kramer-Pesch effect in single-quantum
vortex\cite{KramerPesch}. This behaviour is demonstrated by the
dotted black line in Fig.\ref{Fig:CleanOP}c corresponding to
$T=0.05 T_c$. This effect is featured by all vortices with odd
winding numbers originating from the anomalous energy branch
crossing the Fermi level at $\mu=0$ in the
Eq.(\ref{Volovik-spectr}).

\begin{figure}[h!]
 \centerline{\includegraphics[width=1.0\linewidth]{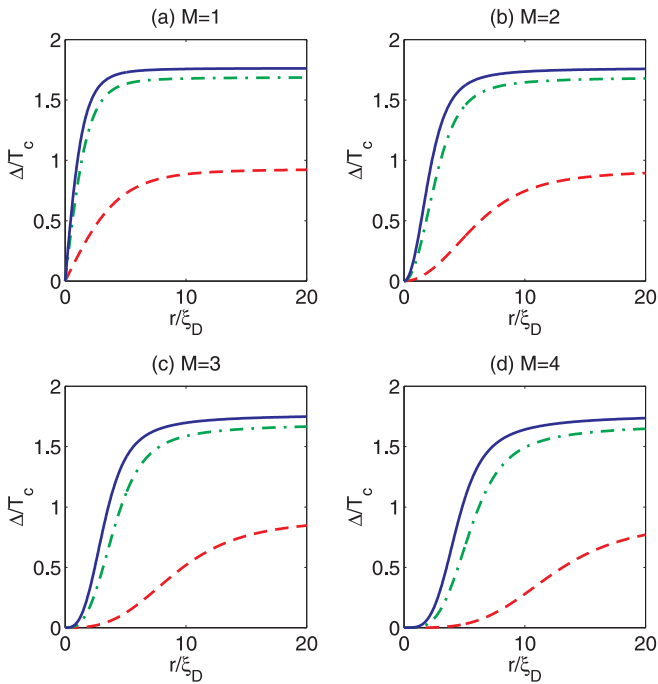}}
 \caption{\label{Fig:DirtyOp}
 The distribution of the order parameter around vortex core in
 dirty superconductor at different temperatures.
 The panels (a,b,c,d) correspond to the winding numbers
 $M=1,2,3,4$ correspondingly. Blue solid, green dash-dotted and red dashed
 lines correspond to the temperature
 $T/T_c=0.1;\;0.5;\; 0.9$. }
 \end{figure}

Next consider the case of dirty superconductor and calculate the
core structures of multiquantum vortices. The results of
calculation are shown in Fig.(\ref{Fig:DirtyOp}) for the winding
numbers $M=1,2,3,4$ and temperatures $T/T_c= 0.1;\;0.5;\;0.9$. As
expected the vortices in dirty regime do not feature singularities
in the order parameter distribution in contrast to the clean case
considered above.

\begin{figure}[h!]
 \centerline{\includegraphics[width=1.0\linewidth]{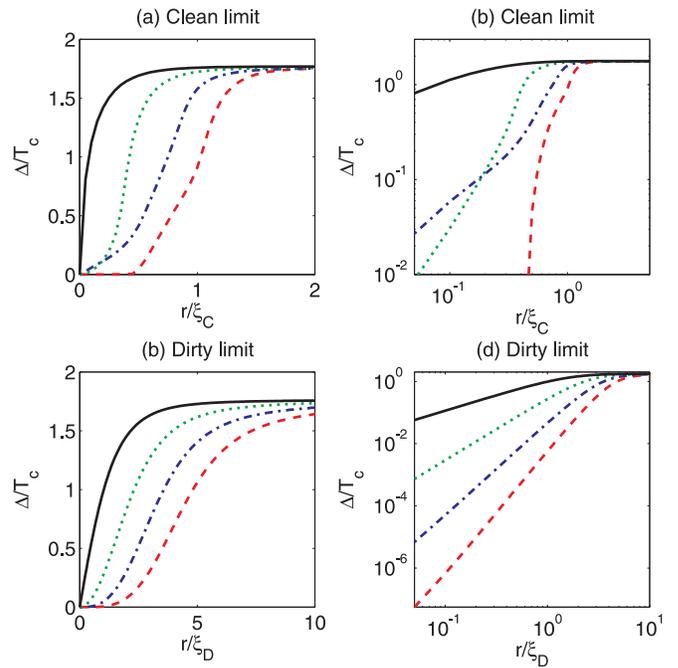}}
 \caption{\label{Fig:CleanDirtyOp}
 The distribution of the order parameter around multiquantum vortex core at $T/T_c=0.1$ in (a) dirty superconductor
 and (b) clean superconductor. Black solid, green dotted, blue dash-dotted and red dashed lines correspond to the
 winding numbers $M=1,2,3,4$. }
 \end{figure}

The comparison of vortex core structures in clean and dirty
superconductors at $T/T_c=0.1$ is presented in
Fig.\ref{Fig:CleanDirtyOp} for the winding numbers $M=1,2,3,4$. To
demonstrate the difference between clean and dirty cases we plot
the dependencies $\Delta=\Delta(r)$ in logarithmic scale in
Figs.\ref{Fig:CleanDirtyOp}(b,d) correspondingly. In the dirty
case the order parameter has Ginzburg-Landau  power law asymptotic
$\Delta(r)= \alpha r^M$ which takes place at $r\rightarrow 0$ even
at very low temperatures $T\ll T_c$. In
Fig.\ref{Fig:CleanDirtyOp}(a,b) the low-temperature behavior
$\Delta (r)$ in the clean case is drastically different from
Ginzdurg-Landau regime. In particular the multiquantum vortex with
$M=3$ shown by blue dash-dotted line in
Fig.\ref{Fig:CleanDirtyOp}a has linear asymptotic at $r=0$. The
slope of linear asymptotic for $M=3$ should grow with decreasing
temperature featuring an analog of Kramer-Pesch effect for
multiquantum vortices. Furthermore the order parameter in $M=4$
vortex shown by red dashed line in Fig.\ref{Fig:CleanDirtyOp}a is
almost zero at finite region $r<R_c$ where $R_c\sim \xi_C/2$. This
behavior is caused by the dominating contribution of the
electronic states corresponding to anomalous branches
(\ref{Volovik-spectr}) to the superconducting order parameter at
$r<R_c$. Thus contribution is zero at $r<min
(\mu_{01},\mu_{02})/k_F$ in the limit $T\rightarrow 0$
\cite{VolovikAnomaly}. Thus the multiquntum vortices with even
winding numbers $M$ are well described by the step-wise vortex
core model used previously for the analytical analysis of the
vortex core spectrum \cite{Melnikov-Vinokur-2002}.

 \section{LDOS profiles of multiquantum vortices}
 \label{ldos}
\begin{figure}[h!]
 \centerline{\includegraphics[width=1.0\linewidth]{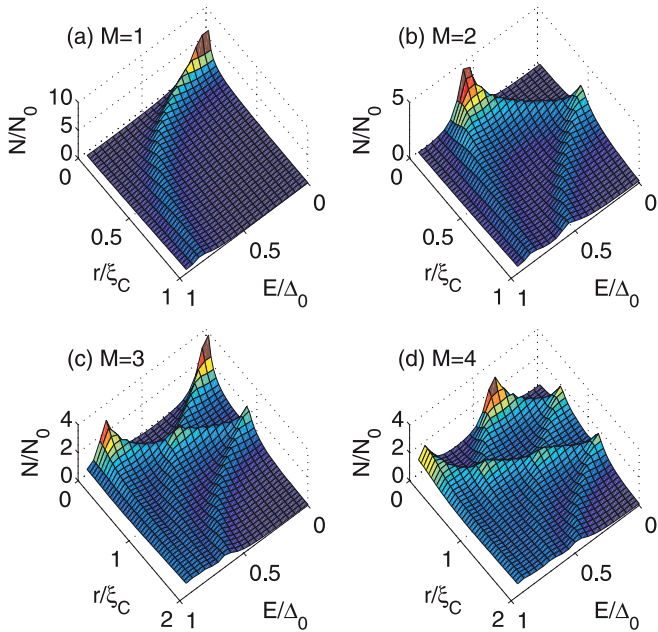}}
 \caption{\label{Fig:CleanLDOS}
 The distribution of the LDOS around  vortex cores at $T/T_c=0.1$ in clean superconductor
 as function of energy and distance from the vortex core $N=N(r,E)$.
 The panels (a,b,c,d) correspond to the values of vorticity
 $M=1,2,3,4$ correspondingly. }
 \end{figure}

 Having in hand the order parameter structures calculated
 self-consistently is Sec.(\ref{op}) we calculate the LDOS distributions
formed by the electronic states localized at the vortex
 core. We start with the case of clean superconductor which is
 known to demonstrate peculiar profiles of LDOS originating from
 multiple anomalous
  energy branches of localized electrons
  \cite{TanakaMultiquantum,SalomaaMultiquantum}. Here we calculate
  the LDOS distributions for the winding numbers $M=1,2,3,4$
  shown in Fig.(\ref{Fig:CleanLDOS}). The LDOS plots are
  similar to that obtained in the framework of Bogolubov- de
  Gennes theory\cite{TanakaMultiquantum,SalomaaMultiquantum}.

 Introducing a polar
coordinate system $(r,\varphi)$  and defining the $z$ projection
of quasiparticle angular momentum through the impact parameter of
quasiclassical trajectory\cite{Silaev2008} $\mu=- [{\bf r},{\bf
k}_F]\cdot{\bf z}_0$ the LDOS inside the singly quantized vortex
core can be found with the help of Eq.(\ref{Eq:SinglyQuantized})
as follows: $N (E,r) =(k_F/2\pi\xi_C)\int_0^{2\pi}\delta[E-\omega
k_Fr\sin(\varphi-\theta_p)]d\theta_p$.
 Here we evaluate the LDOS summing up over the
quasiparticle states at the trajectories characterized by the
  direction of the quasiparticle linear momentum
${\bf k_F}= k_F (\cos\theta_p,\sin\theta_p)$.
 This expression yields a singular
behaviour of zero energy LDOS at $r>r_0$\cite{Ullah, Maki,
IchiokaStar}: $N(E,r)=1/(2\pi\omega \sqrt{r^2-r_0^2}\xi_C)\approx
N_0 \xi_C/\sqrt{r^2-r_0^2}$, where $N_0=(1/2\pi)m/\hbar^2$ is a
normal metal LDOS and $r_0=E/(\omega k_F )$. Thus the LDOS profile
of singly quantized vortex has the ring form with the radius $r_0$
being a function of energy. The dependence $N=N (E,r)$ is shown in
Fig.(\ref{Fig:CleanLDOS})a for a singly quantized vortex.

In multiquantum vortices the spectrum of low energy states
(\ref{Volovik-spectr}) contains several anomalous branches which
intersect the Fermi level and contribute to the low-energy DOS.
The LDOS profile corresponding to the spectrum
(\ref{Volovik-spectr}) consists of a set of axially symmetric ring
structures\cite{Melnikov-Vinokur-2002,TanakaMultiquantum,SalomaaMultiquantum}.
Note that for an even winding number the anomalous branch crossing
the Fermi level at $\mu=0$ (i.e. at zero impact parameter) is
absent and, as a result, the LDOS peak at the vortex center
disappears. Using the same procedure as for the singly quantized
vortices and the spectrum (\ref{Volovik-spectr}) we obtain the
LDOS in the form
 $N(E,r)=\sum_{i=1}^M \vartheta(r-r_{0i})/(2\pi\omega_i \sqrt{r^2-r_{0i}^2}\xi_C)$
 where $r_{0i}=[\mu_{0i}+E/\omega_i]/k_F$ and the step function $\vartheta(r)=0(1)$ at $r<(>)r_{0i}$.
  At $E=0$ the spectrum
 is symmetric so that the LDOS profile has $M/2$ peaks for even $M$
 and $(M+1)/2$ for odd $M$. At $E\neq 0$ the degeneracy is removed
 and each peak splits by two as can be seen from the LDOS plots in
 Fig.(\ref{Fig:CleanLDOS}).

\begin{figure}[h!]
 \centerline{\includegraphics[width=1.0\linewidth]{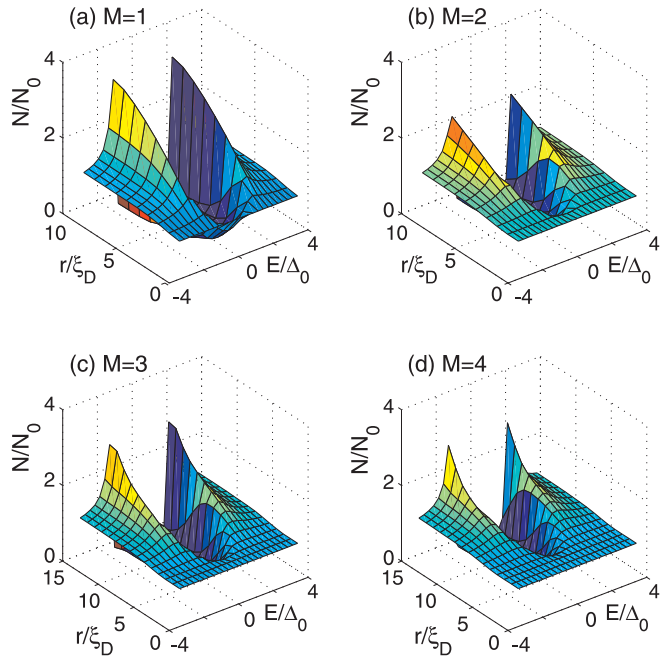}}
 \caption{\label{Fig:DirtyLDOS}
 The distributions of LDOS around vortex cores at $T/T_c=0.1$ in dirty superconductor
  as functions of energy and distance from the vortex core $N=N(r,E)$.
    The panels (a,b,c,d) correspond to the values of winding
    number $M=1,2,3,4$. }
 \end{figure}

\begin{figure}[h!]
 \centerline{\includegraphics[width=1.0\linewidth]{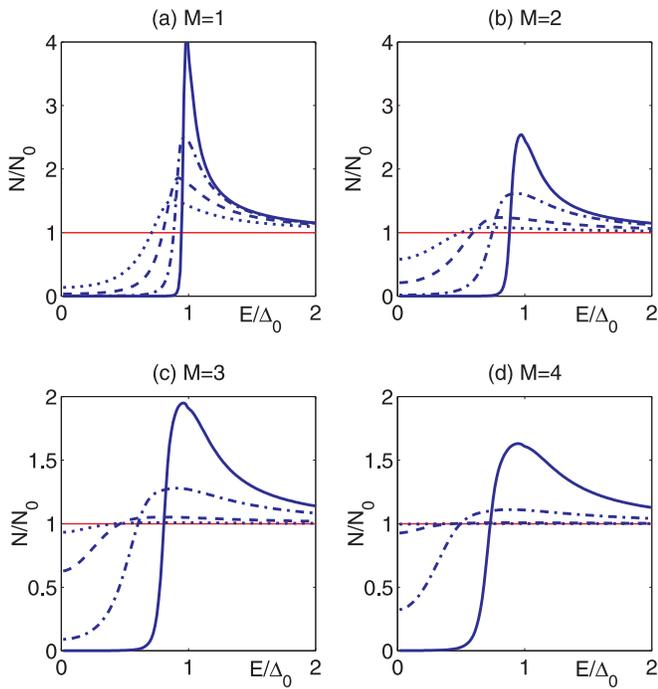}}
 \caption{\label{Fig:DirtyLDOS(E)}
 The cross sections $N=N(E)$ at different values of distance from the vortex center
 $r$ in dirty superconductor at $T/T_c=0.1$. The panels (a,b,c,d) correspond to the values of winding
    number $M=1,2,3,4$. Blue dotted, dash-dotted, dashed and solid lines correspond to the distances $r/\xi_D=2;3;5;10$.
Thin solid red line indicates the flat LDOS at the vortex center
$r=0$.   }
 \end{figure}

Smearing of energy levels due to scattering effects leads to a
reduction of LDOS peak at the vortex center. However, the LDOS
peak survives even in "dirty" limit when a mean free path is
smaller than a coherence length $l<\xi$. To find the form of LDOS
peak at the vortex core we consider the dirty case described by
Usadel Eq.(\ref{Usadel-Reduced}). The LDOS distributions around
multiquantum vortices calculated according to
Eqs.(\ref{Usadel-Reduced},\ref{Eq:DOSdirty}) are shown in
Fig.(\ref{Fig:DirtyLDOS}). The profiles of LDOS at zero energy
level $N=N(r)$ in multiquantum vortices $M>1$ feature plateau near
the vortex center . This is in high contrast to the case of singly
quantized vortex $M=1$. The cross sections $N=N(E)$ at different
values of distance from the vortex center are shown in
Fig.(\ref{Fig:DirtyLDOS(E)}) for $T/T_c=0.1$ and the  winding
numbers $M=1,2,3,4$. These plots clearly demonstrate that with
tunneling spectroscopy measurements it is hard to determine the
center of the multiquantum vortex core for $M>2$. Indeed for $M=3$
the dependencies $N=N(E)$ for $r=0$ and $r=2\xi_D$ are very close
to each other. For $M=4$ the same is true up to $r=3\xi_D$.

\begin{figure}[h!]
 \centerline{\includegraphics[width=1.0\linewidth]{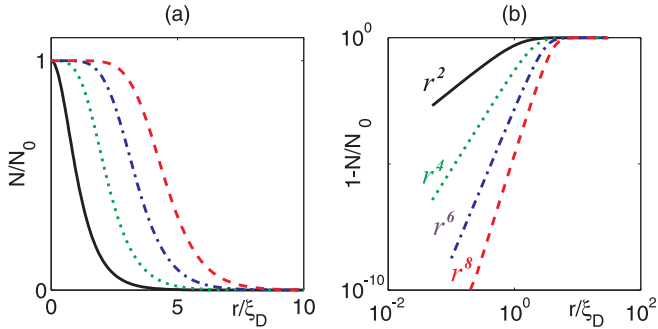}}
 \caption{\label{Fig:DirtyLDOS1D}
 (a) The LDOS profiles for zero energy $E=0$ around vortices at $T/T_c=0.1$ in dirty superconductor
  as function of the distance from the vortex center $N=N(r)$.
  (b) The logarithmic plot of $\delta N (r)=1-N(r)/N_0$ demonstrating the power
 law asymptotic $\delta N (r)\sim r^{2M}$ at $r\rightarrow 0$. Black solid, green dotted, blue dash-dotted and red dashed lines correspond to the
 winding numbers $M=1,2,3,4$. }
 \end{figure}

In fact the discussed LDOS plateau occur due to the very slow
spatial dependence of $\delta N (r) = 1-N(r)/N_0$ at small $r$
which can deduced directly from
Eqs.(\ref{Usadel-Reduced},\ref{Eq:DOSdirty}). Indeed linearizing
the Eq.(\ref{Usadel-Reduced}) for $\omega=0$ we obtain
 \begin{equation}\label{Eq:asymptotic}
 \left[\frac{1}{r}\frac{d}{dr} \left(r \frac{d}{dr} \theta\right) -\frac{M^2}{r^2} + \Delta (r)\right]\theta = 0
 \end{equation}
 which define the asymptotic $\theta (r) = \alpha r^M$.
 Next the Eq. (\ref{Eq:DOSdirty}) yields the LDOS deviation
 $\delta N =  \theta^2/2 =  \alpha^2 r^{2M}/2$. This analytical asymptotic perfectly agrees with the
numerical results which can be seen from the logarithmic scale
plot of $N(r)$ in Fig.(\ref{Fig:DirtyLDOS}) b. An interesting
feature of such LDOS plateau is that they survive at the distances
compared to the size of the multiquantum vortex core which is much
larger than the coherence length $\xi_D$. That is we find that the
size of the plateau shown in Fig.(\ref{Fig:DirtyLDOS1D}) is
approximately given by $R_p= M\xi_D/2$ for $M>1$.

\section{Conclusion}
\label{summary} To summarize we have calculated self-consistently
in the framework of quasiclassical Eilenberger theory the order
parameter structures of multiquantum vortices together with the
local density of electronic states both in clean and dirty
superconductors. We have fond that at the temperatures near $T_c$
the order parameter profiles of vortices are qualitatively similar
in clean and dirty regimes (compare the dependencies $\Delta (r)$
 for $T=0.9 T_c$ shown by red dashed curves in Figs.\ref{Fig:CleanOP} and
\ref{Fig:DirtyOp} for clean and dirty cases correspondingly). In
this temperature regime the order parameter asymptotic at
$r\rightarrow 0$ is determined by the power law $\Delta (r) =
\alpha r^M$ which is consistent with the result of Gizburg-Landau
theory valid at $|T/T_c-1| \ll 1$.

On the other hand in the low temperature limit $T=0.1 T_c$
vortices in clean superconductor demonstrate the anomalies in the
order parameter distribution - the singularities of the derivative
$d\Delta/dr$ predicted in Ref.(\cite{VolovikAnomaly}) and shown in
Fig.(\ref{Fig:VortexCoreAnomaly}). Such singularities occur due to
the contribution of anomalous electronic spectral branches to the
 order parameter. The singular behavior of
$d\Delta/dr$ in multiquantum vortices is analogous to the
Kramer-Pesch effect\cite{KramerPesch} taking place for singly
quantized vortex $M=1$ which has steep order parameter slope
$d\Delta/dr (r=0) \sim 1/T $ at $T\rightarrow 0$.
In dirty superconductors the asymptotic $\Delta(r\rightarrow 0)$
at the vortex core obeys the Ginzburg-Landau power law behavior
even at low temperature $T=0.1 T_c$ which is clearly demonstrated
in logarithmic scale plots in Fig.\ref{Fig:CleanDirtyOp}d.

In the framework of quasiclassical theory we calculated the LDOS
distributions in muliquantum vortices with winding numbers
$M=1,2,3,4$. The LDOS profiles in the clean regime are similar to
that obtained previously with the help of Bogolubov-de Gennes
theory\cite{TanakaMultiquantum,SalomaaMultiquantum}. Most
importantly we determined the LDOS profiles in dirty regime which
directly correspond to the modern experiments on scanning
tunneling microscopy of multiquantum vortices in mesoscopic
superconductors. The zero energy LDOS profile near the vortex
center is shown to be $N(r)/N_0=1-\alpha r^{2M}$ which holds with
good accuracy
 at $r<M \xi_D/2$. Thus for the values of $M>2$ the LDOS profile
 is almost flat at the sizable region near the vortex center $r<M
 \xi_D/2$ (see Fig.\ref{Fig:DirtyLDOS1D}). Such LDOS plateau can be considered as a hallmark of multiquantum
 vortex formation revealed by STM in dirty
 mesoscopic superconductors\cite{RoditchevPRL2011,RoditchevPRL2009}.

\section{Acknowledgements}
 This work was supported, in part by Russian
Foundation for Basic Research Grant N 13-02-01011 and Russian
President Foundation (SP- 6811.2013.5). Discussion with Dr. Vasily
Stolyarov is greatly acknowledged .

\end{document}